\documentclass[osajnl,twocolumn,showpacs, superscriptaddress,10pt]{revtex4-1}

\usepackage[final]{hyperref}
\usepackage{amsmath,amssymb,amsfonts}       
\usepackage{graphicx}
\usepackage{lipsum}
\usepackage{color}

\newcommand{\average}[1]{\left \langle {#1} \right \rangle}

\newcommand{\T}{\mathsf{T}} 

\newcommand{\Proj}{\mathbf{H}_\thetavec} 
\newcommand{\CovMat}{\boldsymbol{\Sigma}} 

\newcommand{\Aar}{\mathbf{A}} 
\newcommand{\Barr}{\mathbf{B}} 
\newcommand{\Car}{\mathbf{C}} 

\newcommand{\D}{\mathbf{G}} 

\newcommand{\tr}{\operatorname{trace}} 

\newcommand{\thetavec}{{\boldsymbol{\theta}}}

\newcommand{\rhovec}{\boldsymbol{\rho}}
\newcommand{\etavec}{{\boldsymbol{\eta}}} 
\newcommand{\phivec}{{\boldsymbol{\phi}}} 
\newcommand{\varphivec}{{\boldsymbol{\varphi}}} 
\newcommand{\varepsilonvec}{{\boldsymbol{\varepsilon}}}

\newcommand{\qed}{\nobreak \ifvmode \relax \else
      \ifdim\lastskip<1.5em \hskip-\lastskip
      \hskip1.5em plus0em minus0.5em \fi \nobreak
      \vrule height0.75em width0.5em depth0.25em\fi}

\begin{document}


\title{Linear prediction of atmospheric wave-fronts for tomographic Adaptive Optics systems: modelling and robustness assessment}


\author{Kate Jackson}\email{Corresponding author: katjac@uvic.ca}
\affiliation{Adaptive Optics Laboratory, University of Victoria, 3800 Finnerty Rd., Victoria, BC, Canada V8P 5C2}

\author{Carlos Correia}
\affiliation{Institute of Astrophysics and Space Sciences, University of Porto, CAUP, Rua das Estrelas, 4150-762 Porto, Portugal}

\author{Olivier Lardi\`ere}
\affiliation{Adaptive Optics Laboratory, University of Victoria, 3800 Finnerty Rd., Victoria, BC, Canada V8P 5C2}

\author{Dave Andersen}
\affiliation{NRC-Herzberg, 5071 West Saanich Rd.,Victoria, BC, Canada}

\author{Colin Bradley}
\affiliation{Adaptive Optics Laboratory, University of Victoria, 3800 Finnerty Rd., Victoria, BC, Canada V8P 5C2}

%

\begin{abstract}
We use a theoretical frame-work to analytically assess temporal prediction error functions on von-K\'arm\'an turbulence when a zonal representation of wave-fronts 
is assumed. Linear prediction models analysed include \textit{auto-regressive} of order up to three, bilinear interpolation functions and a \textit{minimum mean square error} predictor.
This is an extension of the authors' previously published work \cite{correia14} in which the efficacy of various temporal prediction models was established. Here we examine the tolerance of these algorithms to specific forms of model errors, thus defining the expected change in behaviour of the previous results under less ideal conditions. Results show that $\pm 100 \%$ wind-speed error and $\pm 50 \deg$ are tolerable before the best linear predictor delivers poorer performance than the no-prediction case.
 \end{abstract}

\ocis{010.1080, 010.1285}
\maketitle




\noindent Temporal prediction of the atmosphere is a much debated topic. The purpose of prediction is to reduce the error due to servo lag since the turbulence profile changes rapidly (on timescales of a few milliseconds) during the time that it takes to gather sensor information and to compute corrections. 

Unlike single-conjugated AO systems, in tomographic AO direct access to an estimate of the layered wave-front (WF) is provided at the end of the tomographic step. With turbulence estimated in a discrete number of layers, the frozen-flow approximation can now be called upon with a higher degree of fidelity \cite{poyneer09}. 

Temporal prediction is useful for both Multi-Conjugate AO (MCAO) and Multi-Object AO (MOAO) as a means to increase the sky-coverage \cite{correia12, correia14}. Allowing for greater integration times whilst compensating for the lag error by applying a predictive algorithm enables the system to guide on fainter sources \cite{correia14}. Provided information on the dynamics of the atmospheric turbulence is available or can be construed, one should be able to obtain a more accurate estimate of the WF at the time a set of commands is applied to the deformable mirror (DM) and therefore improve performance. Lag errors are also considered a serious limitation in high-contrast imaging systems where the broadening of the PSF along the main axis of wind-blown turbulence severely limits contrast at small separations \cite{kasper12}.

In this letter we discuss alternatives for time-progressing the atmospheric wave-fronts namely a near-Markovian model, auto-regressive models and spatial shifting under frozen-flow \cite{correia14, gavel04, piatrou07}. 
The main goal of our work is to provide insight into the accuracy and robustness bounds of such models in view of a contemporary application to the Raven science and technology demonstrator installed on the Subaru Telescope \cite{andersen12}.

Under the hypothesis that the turbulent atmosphere is a sum of $L$ thin layers located at a discrete number of different altitudes $h_l$, the aperture-plane phase $\phivec(\rhovec,\thetavec,t) $ indexed by the bi-dimensional spatial coordinate vector $\rhovec = [\rho_x, \rho_y]$ in direction $\thetavec = [\theta_x, \theta_y]$ at time $t$ is defined as
 \begin{equation}
\phivec(\rhovec,\thetavec,t) = \Proj \varphivec(\rhovec,t) = \sum_{l=1}^{L} \omega_l \varphivec_l(\rhovec + h_l \thetavec,t) 
\end{equation}
where $\varphivec_l(\rhovec,t)$ is the $l^{th}$-layer wave-front, $\omega_l$ is the $l^{th}$ layer strength and $\Proj$ is a propagation operator in the near-field approximation that relates the aperture-plane WF to the wave-front defined over a discrete number of $L$ layers in the volume by adding and interpolating $\varphivec_l(\rhovec + h_l \thetavec,t)$ on the aperture-plane computational grid.

In the following we assume a point-wise representation of the WF. 
In both MCAO and MOAO, pseudo open-loop or straight open-loop estimation of the phase is performed from noisy measurements involving derived quantities related to $\phivec(\rhovec,\thetavec,t)$ through the operator $\D$ with additive noise $\etavec(\rhovec, t)$ in Eq. (\ref{eq:WFR}) --
by means of a linear wave-front reconstructor. The goal is thus to estimate 
\begin{equation}\label{eq:WFR}
\widehat{\varphivec}(\rhovec,t+\tau) = \mathbf{W} [\D \phivec(\rhovec, t) + \etavec(\rhovec, t)]
\end{equation}
where $\mathbf{W} = \mathbf{W}_t \mathbf{W}_\rho$ is a linear reconstructor that performs spatial (indicated by subscript $\rhovec$) and temporal (indicated by subscript $t$) phase estimation sequentially.

The progression of WFs with time is not a well-defined process. We use the simplifying Taylor or frozen-flow approximation; under this hypothesis,  time-progressing the phase is equivalent to spatially shifting it in both 'x' and 'y' directions by $[\Delta_x, \Delta_y] = [-v_x\tau, -v_y\tau]$, the product of the components of the wind velocity vector, $\mathbf{v} = [v_x; v_y]$ with the lag, $\tau$. Thus the turbulence profile at time $(t + \tau)$ and height $l$ is obtained simply by translating the profile by $\Delta \rhovec = \mathbf{v}\tau$. The temporally shifted phase can then be expressed in terms of the spatial shift, $\varphivec(\boldsymbol{\rho},t+\tau) = \varphivec(\boldsymbol{\rho} - \mathbf{v}\tau,t)$. 
Under this assumption, linear temporal predictive models, $\mathcal{A}(\tau)$, are discussed next.

Suppose the current and future WFs are related by
\begin{equation}
\varphivec(\boldsymbol{\rho},t+\tau) = \mathcal{A}(\tau) \varphivec(\boldsymbol{\rho},t) 
\end{equation}
where $\mathcal{A}(\tau)$ is a block-diagonal linear operator (atmospheric layers are considered independent) which time-progresses the WF. For the sake of compactness the layer dependence is dropped in the following but is subtended.



The linear model that minimizes the mean square prediction residual phase variance is \cite{piatrou07}
\begin{equation}\label{eq:Ammse}
  \mathcal{A}^{*} = \arg\min_{\mathcal{A}^*}\average{\Vert\varphivec(t+\tau) - \mathcal{A}^* \varphivec(t) \Vert^2_{L_2(\Omega)}},
\end{equation}
where $L_2$ stands for Euclidean norm, $\Omega$ is the telescope aperture and $\average{\cdot}$ means ensemble averaging over the turbulence statistics.
It is best to introduce the notation $\varphivec(t+\tau) \triangleq \varphivec_{k+1}$ and $\varphivec(t) \triangleq \varphivec_{k}$ which are the sampled phase vectors with $\tau$ lag in between.

The solution to Eq. (\ref{eq:Ammse}) is readily found to be the \textit{minimum mean square error} (MMSE) estimator
\begin{equation}\label{eq:zonal_A_matrix}
  \mathcal{A}^{*} \triangleq \average{\varphivec_{k+1}\varphivec_k^\T }
  \average{ \varphivec_k \varphivec_k^\T }^{-1},
\end{equation}
where $\varphivec_{k+1,l}(\rhovec) = \varphivec_{k,l}(\rhovec- \mathbf{v}_l\tau)$. This predictor is, in what follows, called the \textit{Spatio-Angular predictor} (SA).
The covariance matrices are computed by sampling the covariance function with distances corresponding to all baselines between phase points in the aperture
\begin{align}\label{eq:cov_fcn_zonal}
C_\phi(\rho) = &\left(\frac{L_o}{r_0}\right)^{5/3}\frac{\Gamma(11/6)}{2^{5/6}
  \pi^{8/3}}\times \dots \nonumber \\
&\left[\frac{24}{5}\Gamma\left(\frac{6}{5}\right)\right]\left(\frac{2
    \pi \rho}{L_0}\right)^{5/6}K_{5/6}\left(\frac{2 \pi \rho}{L_0}\right) 
\end{align}
where $L_0$ is the outer scale of turbulence, $r_0$ Fried's parameter,
$\Gamma$ the 'gamma' function and finally $K_{5/6}$ a modified Bessel
function of the third order. Parameter $\rho = |\boldsymbol{\rho}|$ is the distance between any two
phase points in the bi-dimensional plane.

The operator $\mathcal{A}^*$ is invertible. With the turbulence covariance matrix $\CovMat_\varphivec$ translation invariant, then $\mathcal{A}^*\CovMat_\varphivec \mathcal{A}^{*,\T} = \CovMat_\varphivec$.

Furthermore, Eq. (\ref{eq:zonal_A_matrix}) is a general method to generate and predict phase in a 2D plane (any wind velocity can be used) according to the Markovian model 
\begin{equation}\label{eq:ar1_mmse}
  \varphivec_{k+1} = \mathcal{A}^{*} \varphivec_{k}+ \varepsilonvec_k^*
\end{equation}
where $\varepsilonvec_k^*$ is an excitation noise whose covariance function $\CovMat_\varepsilonvec^*$ is fixed to guarantee proper turbulence statistics from either a Kolmogorov or von-K\'arm\'an model. 
Hence $\CovMat_\varepsilonvec^* = \CovMat_\varphivec - \mathcal{A}^{*}\CovMat_\varphivec \mathcal{A}^{*,\T}$, 
since $\average{\varphivec_{k+1} (\varphivec_{k+1})^\T}=
\mathcal{A}^{*}\CovMat_\varphivec \mathcal{A}^{*,\T}+ \CovMat_\varepsilonvec^*$. Note $\CovMat_\varphivec \triangleq \average{\varphivec_{k+1}
  \varphivec_{k+1}^\T} \triangleq
\average{\varphivec_{k}\varphivec_{k}^\T}$, which, due to
stationarity, loses it temporal dependence.

However, it has previously been pointed out \cite{gavel04} that the phase over the aperture is not Markovian since some information on the wind-blown portion that leaves the telescope aperture is dropped. We will nevertheless use this near-Markovian model since the temporal update rate of the phase estimates is often of the order of, or below, the coherence time of the turbulence.

The near-Markovian model of Eq. (\ref{eq:ar1_mmse}) generalizes the one from \cite{assemat06} for simulating infinitely long, non-stationary phase screens to fractional pixel shifts in a bi-dimensional plane. In Ass\'emat's method, a sub-set of the columns of $\mathcal{A}^{*}$ is used (most commonly 2) to include bounded-region correlations. 

\begin{figure}[htpb]
	\begin{center}
            \includegraphics[width=0.25\textwidth]{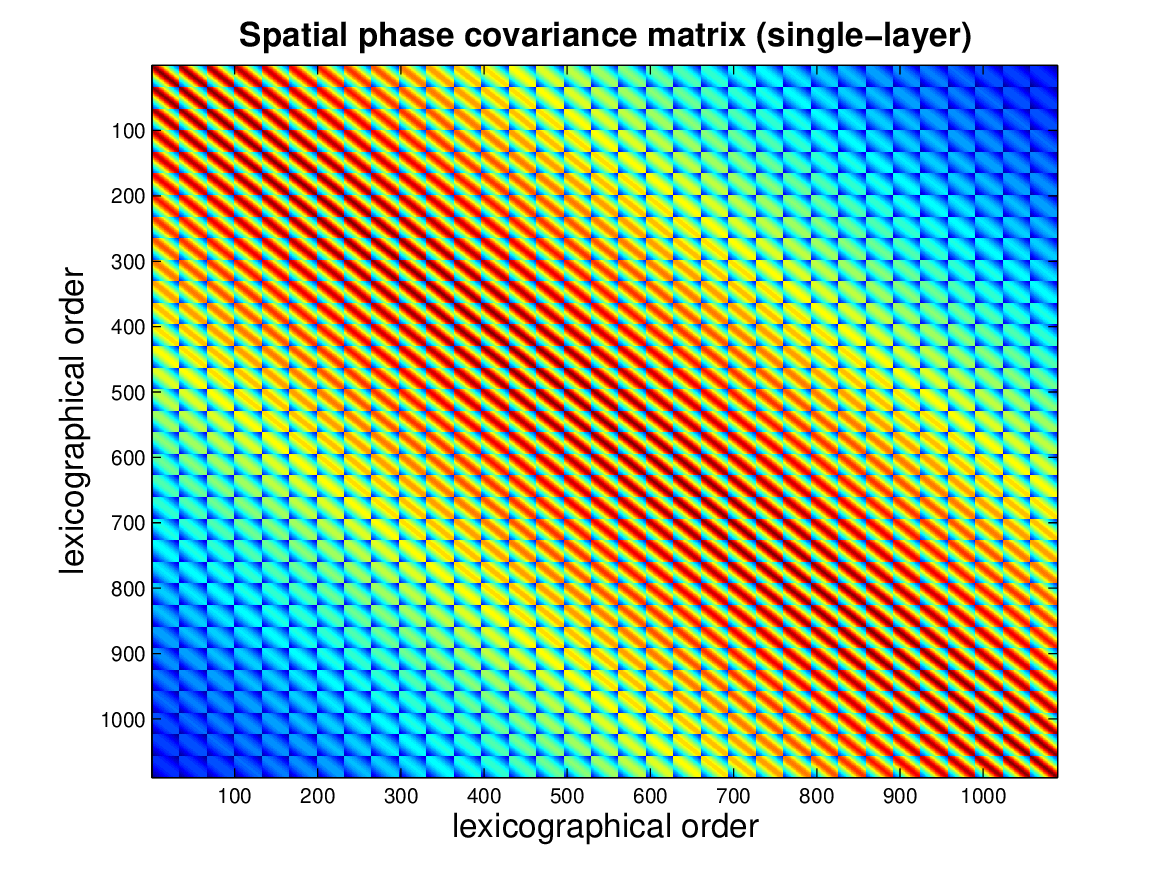}\includegraphics[width=0.25\textwidth]{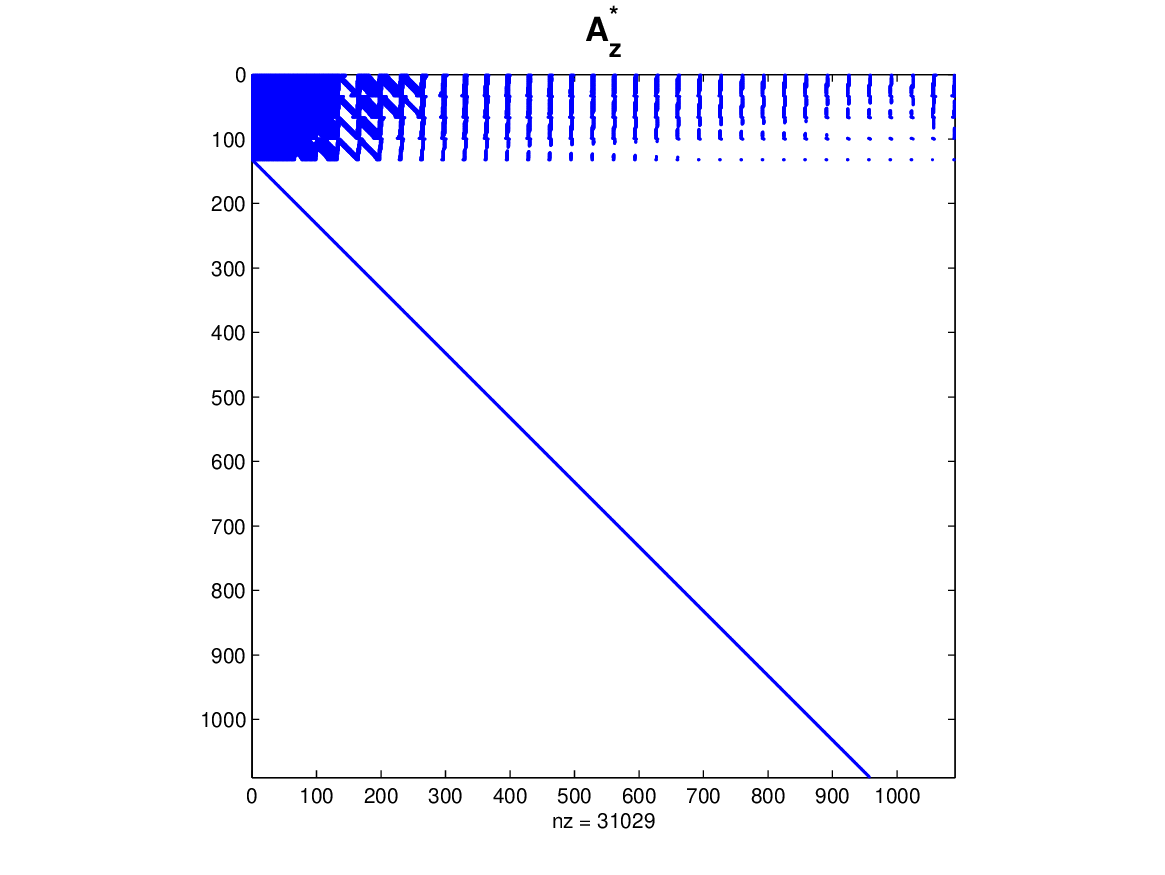}
	\end{center}
	\caption[]
	{\label{fig:Az}
        Left: Spatial phase covariance matrix (single-layer, squared
        pupil, sampled with 33 points across). Right: Matrix $\mathcal{A}^*$ for an horizontal shift of
        0.5m. Sampling 0.25m. $\mathcal{A}^*$ is
        largely dominated by a diagonal term that simply shifts the
        phase in the appropriate direction. The same stands when the
        shifts are not integer factors of the spatial sampling. In
        that case, $\mathcal{A}^*$ becomes more populated with up to
        four main diagonals that interpolate every single point based
        on the neighbouring 4 points. Matrix shown for entries $> 10^{-4}$.
	}
\end{figure}

At this point, two important observations motivate the discussion that follows: i) $\mathcal{A}^{*}$ is strongly diagonally dominated, suggesting that simpler diagonal models, \textit{i.e.} point-wise, could potentially be
applied -- and indeed they have been extensively used in AO simulations in the form of auto-regressive (AR) models \cite{sivo12} albeit with Zernike functions; ii) $\mathcal{A}^{*}$ works out to shift the phase screen in the appropriate wind direction with
shifts given by $|\mathbf{v}_l|cos(\theta_l)$ and $|\mathbf{v}_l|sin(\theta_l)$ for
layer $l = 1\cdots L$. Furthermore, the new turbulence that enters in the
telescope aperture is estimated using spatial correlations with all
the points in the aperture -- see Fig. \ref{fig:Az}. 

As mentioned above, this model is specific for frozen-flow, which is but a part of the actual atmospheric disturbances \cite{poyneer09}. The robustness to parameter mismatches of models with such an underlying assumption must therefore be well understood.

An AR model of order $n$ is defined by the recursion 
\begin{equation}\label{eq:varphi+1}
  \varphivec_{k+1} = f\left(\varphivec_{k}, \cdots
  \varphivec_{k-n-1}\right) + \varepsilonvec_k,
\end{equation}
where $f(\cdots)$ is a linear function yet to be defined and
$\varepsilon_k$ is a Gaussian-distributed, spectrally white, zero-mean random
sequence with covariance given in ref. \cite{correia14}. 

Diagonal AR models are suitable for systems constrained by real-time computational burden. Using simpler diagonal auto-regressive models can be quite appealing as they circumvent $\mathcal{A}^{*}$ being a dense matrix. Although these relatively coarse models are not adapted to simulating atmospheric turbulence, they are used instead for prediction when embedded in the reconstructor and plugged into the dynamical controller, as is done with Kalman filtering, for off-line computation of optimal gains. The AR models are included here because they have been shown to provide stable and improved error reduction within that framework \cite{sivo12, correia12}.


Consider the following models
\begin{align}\label{eq:ar1}
  &\varphivec_{k+1} = \Aar_\text{\tiny{AR1}} \varphivec_{k}+ \varepsilonvec_k^\text{\tiny{AR1}} \\
  &\varphivec_{k+1} = \Aar_\text{\tiny{AR2}}\varphivec_{k} + \Barr_\text{\tiny{AR2}}\varphivec_{k-1}+ \varepsilonvec_k^\text{\tiny{AR2}} \nonumber \\
 &\varphivec_{k+1} = \Aar_\text{\tiny{AR3}}\varphivec_{k} + \Barr_\text{\tiny{AR3}}\varphivec_{k-1}+ \Car_\text{\tiny{AR3}}\varphivec_{k-2} + \varepsilonvec_k^\text{\tiny{AR3}} \nonumber
\end{align}

\begin{figure}[htpb]
	\begin{center}
		\includegraphics[width=0.42\textwidth]{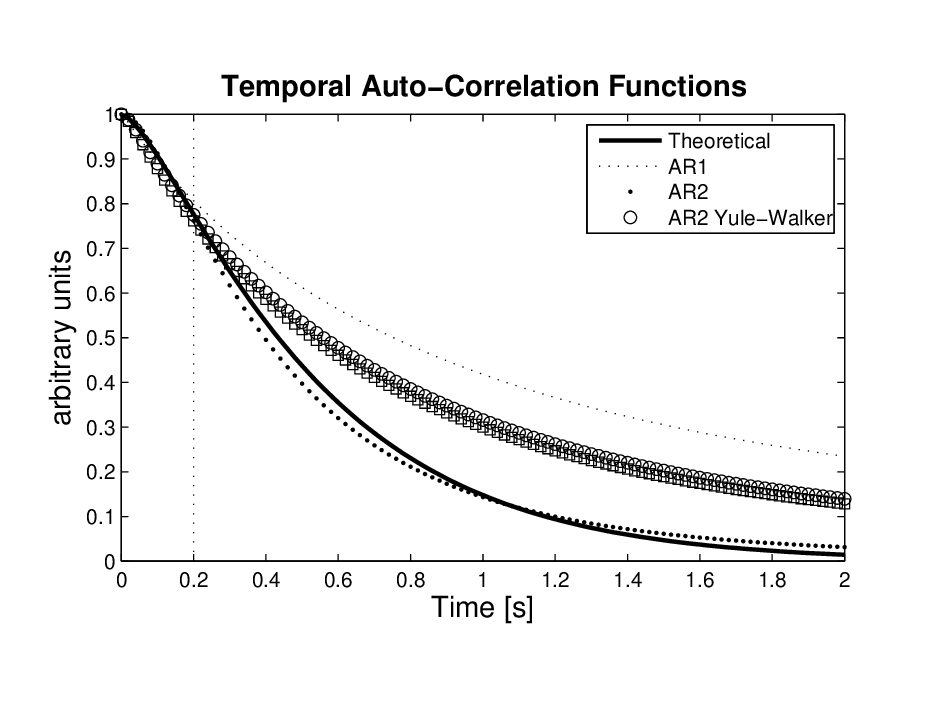}
	\end{center}
	\caption[]
	{\label{fig:TemporalAutoCorrFunctions}
          Left: Temporal auto-correlation functions for turbulence with $r_0=0.155\,m$, $L_0=25\,m$ and identified AR1, AR2 and AR3 models. The vertical dotted line indicates the fitting section.} 
\end{figure}

Figure \ref{fig:TemporalAutoCorrFunctions} depicts the temporal auto-correlation functions for the AR models with the fitted second order model achieving an overall best fit. 

A practical alternative to model fitting is the use of the Yule-Walker equations for auto-regressive model computation to identify ARn models from a discretized covariance function at the AO system's sampling frequency -- also shown in Fig. \ref{fig:TemporalAutoCorrFunctions}.

Figure \ref{fig:Az} depicts an analytical covariance matrix and a thresholded Spatio-Angular prediction matirx, $\mathcal{A}^*$ from
Eq. \eqref{eq:zonal_A_matrix}. It is apparent that the optimal one-step estimator is
largely dominated by a diagonal term that simply shifts the
phase in the appropriate direction. Note however that points outside the pupil entering the telescope (upper band in Fig.\ref{fig:Az}-right) are computed using their respective covariance function with points inside the pupil. We thus expect the bilinear interpolation to be a reasonable approximation to the Spatio-Angular model but with degraded performance. It will be tested and assessed in numerical simulations.

Consider the general formulation for the temporal lag error that is compatible with the case of atmospheric prediction \cite{correia14}
\begin{equation}\label{eq:lagerrordef}
 \sigma^2(\tau) =  \average{\left\Vert\Proj \left(\varphivec_{k+1} -\widehat{\varphivec}_{k+1}\right)\right\Vert^2_{L_2(\Omega)}}
\end{equation}
where $\widehat{\varphivec}_k$ is a layered phase estimate at time step $k$.

In the \textbf{no} prediction case, the estimated phase is simply a replication of the phase at the previous time step, $\widehat{\varphivec}_k = \varphivec_{k-1}$. The temporal lag error from Eq.(\ref{eq:lagerrordef})  becomes
\begin{equation}
  \sigma_0^2(\tau) = \tr\left\{\Proj \left(\CovMat_\varphi-\CovMat_\varphi'\right)\Proj^\T\right\}  \triangleq D_t(\tau)
\end{equation}
which is simply the temporal structure function of phase, with the 1-step covariance matrix $\CovMat_\varphivec'=\boldsymbol{\Sigma_\varphivec}(\rho' = \rho + |\mathbf{v}|\tau)$ computed from Eq. (\ref{eq:cov_fcn_zonal}).

These temporal structure functions can now be expanded for the case
of predicted phase. 
Developing Eq. (\ref{eq:lagerrordef}) using the measurement models given in Eq. \ref{eq:ar1} yields,
\begin{enumerate}
\item First order models (AR1 or SA)
\begin{equation}
\sigma^2_{p=1}(\tau)  = \tr\left\{\Proj \left(\CovMat_\varphi +
    \Aar\CovMat_\varphi\Aar^\T -2 \CovMat_\varphi' \Aar^\T\right)\Proj^\T\right\} , 
\end{equation}
\item Second order models (AR2)
\begin{align}
 \sigma^2_{p=2}(\tau) & = \tr\left\{\Proj \left(\CovMat_\varphi +
    \Aar\CovMat_\varphi\Aar + \Barr\CovMat_\varphi\Barr \right. \right.\nonumber \\ 
& \hspace{10pt}
\left. \left. -2 \Aar \CovMat_\varphi' + 2 \Aar \CovMat_\varphi'\Barr^\T
- 2\Barr \CovMat_\varphi''\right)\Proj^\T\right\}
\end{align}
with $\boldsymbol{\Sigma}_\varphivec''= \boldsymbol{\Sigma_\varphivec}(\rho'' = \rho + 2|\mathbf{v}|\tau)$.

\item Third order models (AR3) 
\begin{align}
  \sigma^2_{p=3}(\tau) & = \text{trace}\left\{\Proj \left(
\CovMat_\varphi +    \Aar\CovMat_\varphi\Aar +
\Barr\CovMat_\varphi\Barr + \right. \right. \nonumber \\ & \hspace{10pt}
\left. \left. \Car\CovMat_\varphi\Car
-2 \Aar \CovMat_\varphi' - 2\Barr \CovMat_\varphi'' - 2\Car \CovMat_\varphi''' \right. \right.\nonumber \\ & \hspace{10pt}
\left. \left.
+ 2 \Aar \CovMat_\varphi' \Barr^\T + 2 \Barr \CovMat_\varphi' \Car^\T + 2 \Aar \CovMat_\varphi'' \Car^\T\right)\Proj^\T
\right\}
\end{align}
with $\boldsymbol{\Sigma}_\varphivec'''= \boldsymbol{\Sigma_\varphivec}(\rho''' = \rho + 3|\mathbf{v}|\tau)$.
\end{enumerate}

\begin{figure}[htpb]
	\begin{center}
		\includegraphics[width=0.39\textwidth]{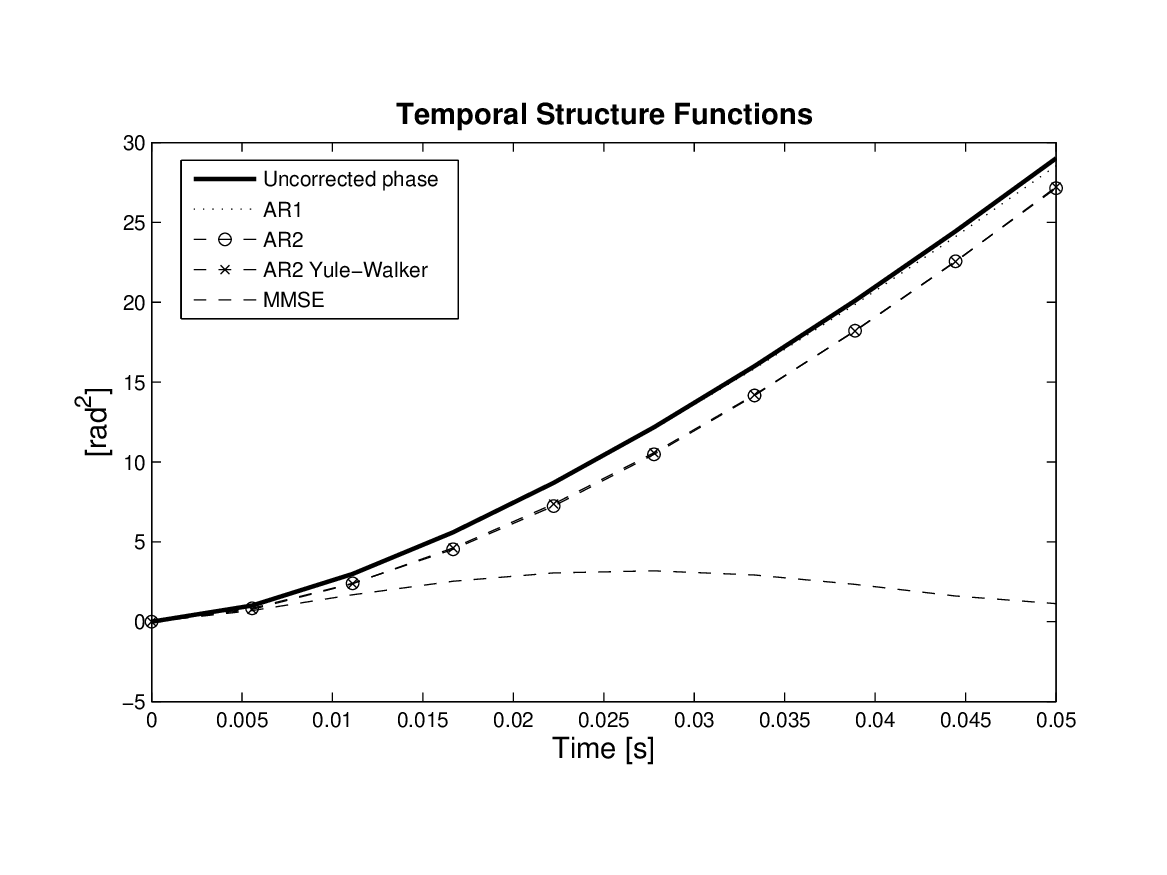} 
	\end{center}
	\caption[]
	{\label{fig:TemporalStructFunct}
	Theoretical temporal structure functions quoted in $rad^2@0.5\mu m$ under same conditions as Fig. \ref{fig:TemporalAutoCorrFunctions}. }
\end{figure}

These equations are revised from those given in \cite{correia14}, which contain a minor transcription error; the numerical results were not affected. The AR models perform very poorly when prediction is considered; in Fig. \ref{fig:TemporalStructFunct} the theoretical temporal structure functions indicate that little improvement is gained from the no-prediction case when the AR models are considered. The SA model performs by far best as previously noted in \cite{correia14}.



\begin{figure}[htpb]
	\begin{center}
		\includegraphics[width=0.37\textwidth]{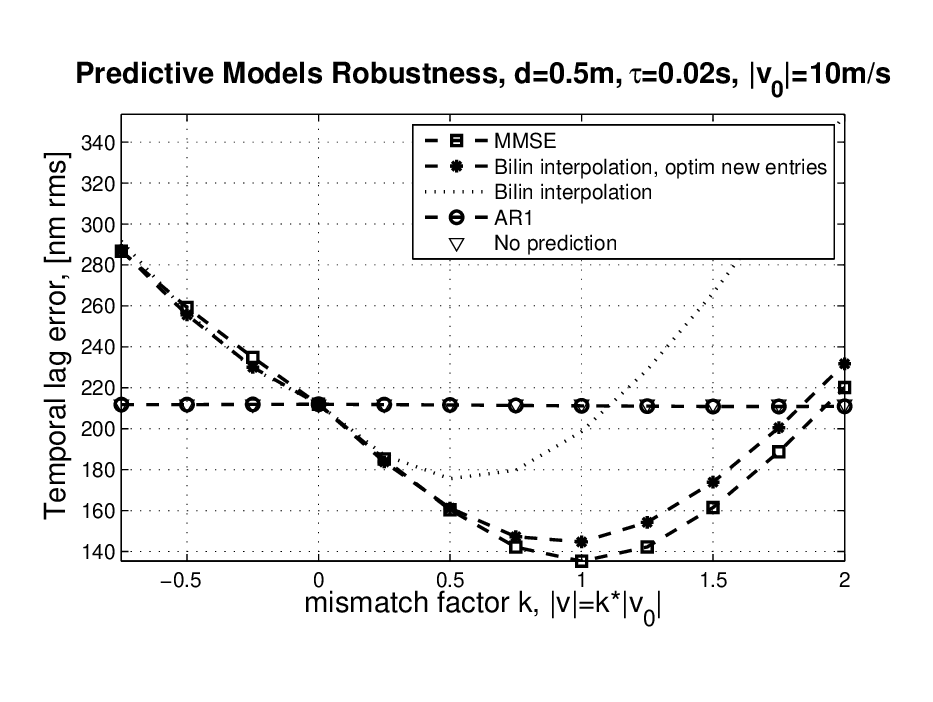}\\\includegraphics[width=0.37\textwidth]{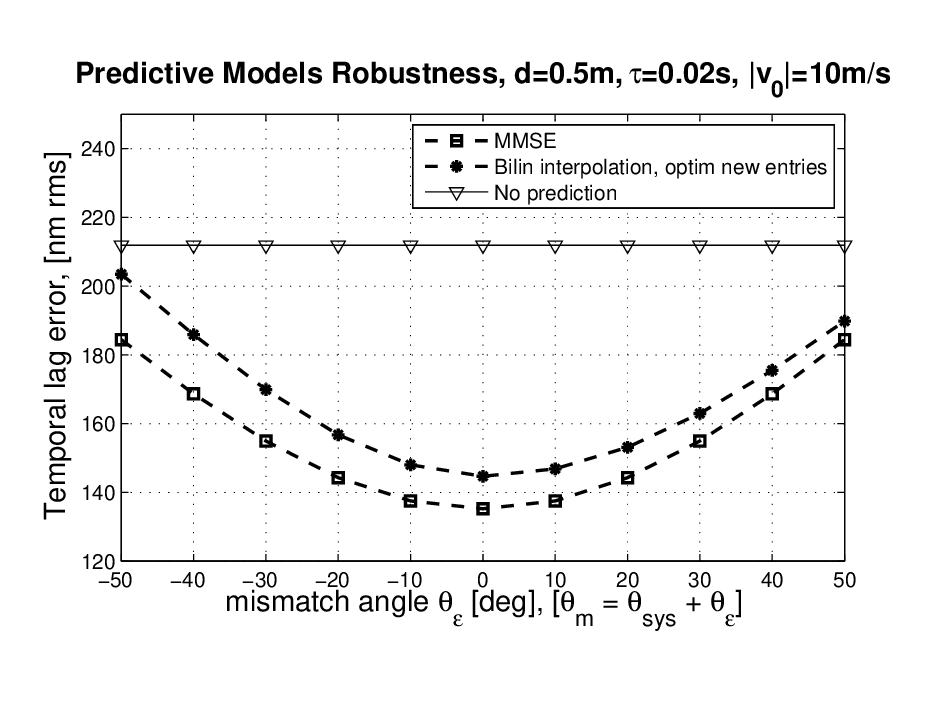}  
	\end{center}
	\caption{(Top:) Robustness wrt wind-speed. (Bottom:) Robustness wrt wind direction.}
	\label{fig:robustness_vs_windspeed}	 
\end{figure}

With the expectation that it is a reasonable approximation to the SA model, we built the bilinear spline interpolation model with two versions. In the first, new points entering the aperture are interpolated with no further information from within the pupil, thus assuming zeroed wave-front outside the pupil. In the second, the new rows are estimated from a correlation function, \textit{i.e.} those new rows in $\mathcal{A}$ (Fig. \ref{fig:Az}) are kept, the remainder being the two-dimensional bilinear spline weights. The results in Fig. \ref{fig:robustness_vs_windspeed} (top) show that if one does not estimate the new points from a correlation function, the loss in performance is quite drastic, with the bilinear spline interpolator achieving best performance for a wind-speed model that is $1/2$ the true wind-speed. When the values of the new points are computed keeping the corresponding columns of $\mathcal{A}^*$ then the performance enhancement is dramatic with a difference of some tens of nm rms. 

Also shown in Fig \ref{fig:robustness_vs_windspeed} are the analytical temporal error functions for a model mismatch in terms of wind-speed and wind direction for the AR1 model and for the full MMSE predictor. 
Results for the AR1 model show that, although it will never reduce temporal lag error, it does not increase it either even in very poorly identified conditions. Results for the MMSE model (and similarly for the bilinear interpolation with optimized new entries into the pupil) indicate that the wind speed of a given layer can be over estimated by up to a factor of 2 before lag error reaches the equivalent of no prediction. An error in wind direction of up to $50^o$ still reduces temporal lag error; results are given for a 50Hz sample rate, but the bounds hold in the range of 10Hz - 1kHz.



\bigskip

The authors acknowledge the support by the European Research Council/European Community under the FP7 through Starting Grant agreement number 239953 and the Marie Curie Intra-European Fellowship grant with reference FP7-PEOPLE-2011-IEF, number 300162.

\end{document}